\documentclass[a4paper]{jpconf}
\usepackage{graphicx}
\begin{document}
\title{Supernova neutrino detection}

\author{Kate Scholberg}

\address{Department of Physics, Duke University, Durham, NC, 27708, USA}

\ead{schol@phy.duke.edu}

\begin{abstract}
The gravitational core collapse of a star produces a huge burst of neutrinos of
all flavors.
A number of detectors worldwide are sensitive to such a burst;
its detection would yield information about both particle physics
and astrophysics.
Sensitivity to all flavors, and ability to
tag different interactions, will be key for extraction of
information.  Here I will survey the capabilities of current
and future detectors for detection of supernova neutrinos from the Milky Way and beyond.
\end{abstract}

\section{The supernova neutrino signal}

The gravitational collapse of the core of a massive stars
entails a vast release of energy.  Because particles with only
weak interactions can readily
escape the
star on a timescale of tens of seconds, an
overwhelming fraction of the binding energy is carried away by neutrinos.
The neutrino burst from a Galactic supernova
can be detected in terrestrial detectors.
As of this writing, the only supernova for which neutrinos have been
detected is SN1987A, for which a total of 19 neutrinos were observed
in two water Cherenkov detectors\cite{Bionta:1987qt,Hirata:1987hu}; 
scintillation
detectors\cite{Alekseev:1987ej,Aglietta:1987it} also reported observations.  These observations
confirmed the expected general features of gravitational collapse,
but the data were insufficient to distinguish fine details of
different models.

The baseline model predicts a burst of neutrinos of total energy
a few times 10$^{53}$ ergs.  The expected proto-neutron star core temperature corresponds
to neutrino energies in the few to tens of MeV range.  In the most
straightforward picture, one expects $\langle
E_{\nu_{\mu,\tau}}\rangle > \langle E_{\bar{\nu}_e} \rangle > \langle
E_{\nu_e} \rangle$, because neutrino species with fewer interactions
with the core's matter
will emerge from deeper, and hence hotter regions of the star.
However, some recent studies (\textit{e.g.}\cite{Keil:2002in})
bring into question the robustness of this
prediction, since scattering may degrade this hierarchy of energies.
The timescale of the burst is tens of seconds (consistent with the 1987A
measurements), with a higher rate during the first few seconds.  The
neutrinos will emerge from the collapsed core well before any
supernova photons.  Possibly, the flux could be modulated by formation
of a black hole or other events early in the neutron star's life.  The
neutrino burst includes all flavors of $\nu$ and $\bar{\nu}$, and the
generic expectation is for the neutrino energy to be roughly
equipartitioned among the different flavors.

\section{What we can learn}

A nearby core collapse supernova would be a neutrino experimentalist's
dream, as well as an astrophysicist's.  The huge burst will certainly
help us learn about the core collapse process itself.  The neutrino
burst's time,
flavor and energy structure will bring information about
the explosion mechanism, accretion, possible quark matter 
or black hole formation, and so on.  In addition, we can learn about
neutrinos themselves.  For instance, absolute
neutrino mass leads to an energy-dependent time of flight delay
as neutrinos travel from their source to Earth; however,
it will be difficult to improve on current laboratory limits.  The
parameters governing neutrino oscillations will imprint themselves on
the neutrino signal.  As the neutrinos propagate through the stellar
matter, they may undergo MSW-type resonance transitions in regions of
specific matter density; in particular, there may be signatures of the
unknown mixing angle $\theta_{13}$ and neutrino mass hierarchy
(\textit{e.g.} \cite{Dighe:1999bi}).  Other properties of neutrinos
may also yield interesting effects, as will various proposed exotic
physics scenarios: in particular, the observed cooling timescale
allows one to set limits on coupling to axions, large extra dimensions,
and other exotic physics (\textit{e.g.}\cite{Raffelt:1997ac,Hannestad:2001jv}),
since any large coupling would allow extra energy to escape from
the star, and lead to a neutrino signal which is abbreviated with respect to
the expected burst.  Again, the measured time, flavor and energy structure of
the burst will contain the signatures of unknown physics.

However, a difficulty here is that both core collapse physics and neutrino
physics affect the nature of the burst, and it may not be trivial to
disentangle the two.  To learn about neutrinos, one must make
assumptions about the nature of the collapse, and vice versa.  
Nevertheless, some
features of the collapse are more robust than others, allowing
model-independent studies.  Also, one may 
cancel supernova model-dependent uncertainties 
in the study of neutrino oscillations 
by comparing fluxes
measured at different locations on the Earth, and one may even look
for matter-induced oscillation features in the spectrum of neutrinos measured
in a single detector
\cite{Lunardini:2001pb,Dighe:2003jg}.  Clearly, the more
information we can gather about the flavor, energy and time structure
of the burst, in as many detectors around the world as possible,
the better chance we will have of disentangling the various
effects.

One other potential scientific gift from a neutrino burst is an early
warning of a supernova's occurrence: the neutrinos emerge promptly
from the dense core, while astronomers must wait hours for the first
photons to appear as the shock wave emerges from the stellar envelope.
The SNEWS\cite{Antonioli:2004zb} network exists to provide such an
early warning to astronomers (and others), which may allow
observations from the very early (and previously rarely-observed)
turn-on of the supernova light.  Clearly, the more information that
can be gathered, in all wavelengths (and also perhaps in gravitational
waves), the better.  Because core collapses are rare events (a few per
century), it is essential to be prepared.

\section{Detector technologies}

From a neutrino experimentalist's point of view, the basic strategy is
to prepare to collect as many neutrino events as possible, of as many
flavors as possible.  A back of the envelope calculation shows that
one typically gets a few hundred neutrino interactions per kton of
detector material for a core collapse event at the center of the Milky
Way, 8.5~kpc away.  For a successful observation, the detector
background rate must not exceed the supernova signal rate in a 10
second burst: this criterion is easy to satisfy for underground
detectors, and is even thinkable for many near-surface detectors\cite{Sharp:2002as}.  One
would like to have event-by-event timing resolution, ability to
measure neutrino energies, and if possible, ability to use the
neutrino information to point back to the supernova.  Sensitivity to
all flavors of the burst is extremely desirable: $\nu_\mu$ and
$\nu_\tau$ flavors comprise two-thirds of the burst's energy, but
because supernova neutrino energies rarely exceed a few tens
of MeV, these components of the flux are overwhelmingly below charged
current (CC) interaction threshold, and neutral current (NC)
sensitivity is required to detect them.  As a final point, it will be
especially valuable for detectors to have \textit{ability to tag}
interactions as $\nu_e$, $\bar{\nu}_e$, and $\nu_{\mu,\tau}$ as well
as just to collect them.

\subsection{Inverse beta decay:}
Currently the world's primary sensitivity to supernova neutrinos
is via that old workhorse of neutrino physics, inverse beta decay:
$\bar{\nu}_e+p\rightarrow e^++n$.
In this reaction, the produced positron has the energy of the neutrino, less
1.8~MeV; the positron's energy loss is the primary means of
detection.  There are furthermore two possible tags of inverse beta
decay: a prompt positron annihilation produces two 0.511~MeV $\gamma$
rays, and the neutron may also be observable via its time-delayed capture on
a nucleus.  Capture of a neutron by a free proton produces a 180
$\mu$s-delayed 2.2~MeV $\gamma$ ray.

In any detector with lots of free protons, inverse beta decay
typically dominates by orders of magnitude.  The reaction has a mild
energy-dependent anisotropy\cite{Vogel:1999zy}.  Examples of detectors leaning
heavily on this reaction are hydrocarbon-based, and usually scintillating
(e.g. Baksan,\cite{Alekseev:2002ji,Alekseev:1998ib}, LVD\cite{Aglietta:2003gi,Aglietta:1992dy}, KamLAND\cite{KTolich, Decowski}, 
Borexino\cite{Monzani:2006jg,Cadonati:2000kq},
and mini-BooNE\cite{Sharp:2002as}.)  Scintillation detectors can often
achieve quite low (sub-MeV) energy thresholds, and therefore have
potential for neutron capture and/or $\gamma$ tagging. However because
scintillation light is emitted isotropically, pointing capability is
generally poor.

Water Cherenkov detectors (e.g. Super-Kamiokande\cite{Namba:2003gc,Fukuda:2002uc}) 
also have a
high rate of inverse beta decays, but have difficulty with tagging
neutrons due to high energy thresholds.  A recent suggestion
to spike water with gadolinium trichloride\cite{Beacom:2003nk} which
may allow tagging.  Gd has a huge neutron capture cross-section; the
resulting $\gamma$-rays can then be observed via the Cherenkov
radiation from Compton scatters.  The Gd-capture technique has been
successfully used in small scintillation detectors (\textit{e.g.} CHOOZ\cite{Apollonio:1999jg})
and is currently under study for Super-Kamiokande.

The use of water Cherenkov detectors for supernova neutrino detection
can be extended to detectors like AMANDA/IceCube that are made of long
strings of sparsely distributed photomultiplier tubes embedded in ice
or water.  Such sparse PMT array detectors are nominally high energy
($>$~GeV) neutrino detectors.  They cannot record MeV neutrinos on an
event-by-event basis; nevertheless they may be able to observe a
coincident increase in single count rates from many phototubes due to a
large number of inverse-beta-decay-induced Cherenkov photons in the
surrounding ice or water\cite{Halzen:1994xe,Halzen:1995ex,Ahrens:2001tz}.

\subsection{Other charged current reactions:}
Charged current interactions can occur for bound as well as free nucleons.
Reactions of both $\nu_e$ and $\bar{\nu}_e$ can occur, with the production of
an electron or positron:
$\nu_e+(N,Z)\rightarrow(N-1,Z+1)+e^-; 
\bar{\nu}_e+(N,Z)\rightarrow(N+1,Z-1)+e^+.$

Cross-sections are typically smaller for bound than for free nucleons, but
can nevertheless be non-negligible.  The charged lepton is usually
observable, and CC interactions sometimes can be tagged in other ways,
\textit{e.g.} via detection of ejected nucleons 
or nuclear de-excitation $\gamma$
rays.  CC cross-sections and the nature of
the observables are dependent on the nuclear physics of the specific
nucleus involved, and in many cases there are large
uncertainties.  Examples of CC interactions useful for supernova
neutrino detection are NC breakup in heavy water
($\nu_e+d\rightarrow p+p+e^-$,
$\bar{\nu}_e+d\rightarrow n+ n +e^+$), interactions with oxygen in
water, ($\nu_e + ^{16,18}\rm{O} \rightarrow ^{16,18}{\rm F} +e^-$,
$\bar{\nu}_e + ^{16}{\rm O} \rightarrow ^{16}{\rm N}+e^+$), 
and interactions with
carbon in scintillator ($\nu_e + ^{12}{\rm C} \rightarrow ^{12}{\rm N} +e^-$,
$\bar{\nu}_e + ^{12}{\rm C}\rightarrow ^{12}{\rm B} +e^+$).  Interactions with
heavier nuclei may also yield high rates: for example,
various lead-based detectors have been proposed (OMNIS, LAND/HALO)\cite{Boyd:2002cq,Hargrove:1996zv}.  A
particularly nice tagged $\nu_e$ channel is available in argon,
$\nu_e+^{40}{\rm Ar}\rightarrow e^{-}+^{40}{\rm K}^{*}$; the $^{40}{\rm K}^{*}$
de-excitation $\gamma$'s would be observable in various proposed large
liquid argon detectors (Icarus, LANNDD)\cite{Bueno:2003ei,Cline:2006st}.  Finally, radiochemical
detectors based on Ga, Cl, and other isotopes could potentially yield
excess events (although without time resolution);
and it is in principle possible
to run some of these in a quasi-real-time mode.

\subsection{Elastic scattering:} Elastic neutrino-electron scattering (ES),
$\nu_{e,x}+e^{-}\rightarrow\nu_{e,x}+e^{-}$, which occurs via both CC
and NC channels, has a relatively small cross-section: the rate is a
few percent of the inverse beta decay rate in scintillator and water
Cherenkov detectors.  Nevertheless the ES component of the supernova
neutrino signal will be especially interesting, because it is
\textit{directional}: the electrons get kicked forward by the
neutrinos with an average angle of about 25$^\circ$.  If the direction
of the kicked electron can be determined (e.g. from a Cherenkov cone),
elastic scattering can be used to learn the location of the supernova, and 
is in
fact the best way of using a neutrino detector to point back to the
supernova's location\cite{Beacom:1998fj}.

\subsection{Neutral current reactions:} Only the $\nu_e$ and $\bar{\nu}_e$ components of the 
supernova neutrino signal are accessible via CC interactions.  Because
NC interactions are flavor-blind, they measure the total flux,
including the $\nu_{\mu}$ and $\nu_{\tau}$ components.  Various NC
interactions on nuclei have cross-sections that yield reasonable
rates, and as for the CC case, sometimes a nice tag is possible via
ejected nucleons or de-excitation $\gamma$'s.  For example, a 15.5~MeV
de-excitation $\gamma$-ray tags the NC excitation of $^{12}{\rm C}^{*}$,
$\nu_x + ^{12}{\rm C} \rightarrow~\nu_x + ^{12}{\rm C}^{*}$; a cascade of
5-10~MeV de-excitation $\gamma$s may also tag
$\nu_x + ^{16}{\rm O} \rightarrow \nu_x + ^{12}{\rm O}^{*}$ in a water
Cherenkov detector
%which will make up $\sim$8\% of the total signal
\cite{Kolbe:2002gk}.

A particularly promising future possibility for NC supernova
neutrino detection is a lead-based neutrino detector, for which the
cross-section is high for NC as well as CC channels. For
$\nu_x + ^{210}{\rm Pb} \rightarrow \nu_x + ^{210}{\rm Pb}^{*}$, the lead
nucleus subsequently emits a neutron.  The one-neutron emission
channel is expected to be predominant for
NC\cite{Fuller:1998kb,Engel:2002hg}, in contrast to a high rate of
double-neutron emission for the CC reaction.  The relative rates for
the different channels in lead depend on neutrino energy, which
promises some spectral information and hence sensitivity to
oscillation effects.  There have been proposals to employ metallic
lead and lead in form of perchlorate. 
A promising recent proposal is HALO\cite{halo}, which 
plans to make use the $^3$He NCD counters from SNO when SNO shuts down
at the end of 2006.  As for the CC $\nu$-nucleus reactions, here again
rates and signatures depend on specific nuclear physics.

Another NC channel which has been not been explored until fairly
recently is neutrino-proton NC elastic scattering,
$\nu + p \rightarrow \nu +p$\cite{Beacom:2002hs}.  The rate is relatively
high, but because the free proton target is heavy, recoil kinetic
energies are low.  The recoils
may nevertheless be observable in large low threshold
scintillation detectors, e.g. KamLAND, even after accounting for
``quenching'' in scintillator.  Neutral current coherent
neutrino-nucleus elastic scattering, $\nu + A \rightarrow \nu +A$,
occurs at even higher rates than $\nu p$ scattering, but because the
targets are yet heavier, recoil energies are yet tinier-- in the tens
of keV range.  This might seem a hopeless situation, but such tiny
recoils are within the reach of novel detectors developed for pp solar
neutrinos or WIMP detection\cite{Horowitz:2003cz}.  For example, a detector 
like CLEAN\cite{McKinsey:2004rk}, which can potentially expand to a 10 ton 
scale,
would observe a few events per ton from an 8.5~kpc
supernova.

\subsection{Detector summary:}
Current and proposed detectors are summarized in Table~\ref{tab:detectors}.
The numbers of events given for a Galactic center supernova should
be taken as uncertain by at least 50\%; not only are there
uncertainties in the collapse models, in many cases the numbers
of observable events depend on assumed thresholds, efficiencies, enrichment,
and other detector-configuration-specific properties.

\begin{table}
\caption{\label{tab:detectors}Summary of current and proposed detectors.}
\begin{center}
\begin{tabular}{llllll}
\br
Detector&Type &Mass (kton) &Location & Events at 8.5~kpc & Status\\ \br
Super-K\cite{Namba:2003gc} & H$_2$O& 32 & Japan& 7000& Running\\
SNO\cite{Virtue:2001mz} & D$_2$O & 1 (D$_2$O) & Canada & 400& Running until\\
    &             & 1.4 (H$_2$O) & & 450& end 2006\\
LVD\cite{Aglietta:1992dy} & C$_n$H$_{2n}$& 1 & Italy& 200& Running\\
KamLAND\cite{KTolich} & C$_n$H$_{2n}$& 1 & Japan& 300& Running\\
Borexino\cite{Monzani:2006jg} & C$_n$H$_{2n}$& 0.3 & Italy& 100& 200x\\
Baksan\cite{Alekseev:1998ib} & C$_n$H$_{2n}$& 0.33 & Russia& 50& Running\\
Mini-BooNE\cite{Sharp:2002as} & C$_n$H$_{2n}$& 0.7 & USA & 200& Running\\
AMANDA/ & Long string& 0.4/PMT & South Pole & N/A& Running\\
IceCube\cite{Ahrens:2001tz} & & &  && Running\\
SAGE\cite{Abdurashitov:1994bc} & Ga & Russia & 0.06 & few & Running\\
Icarus\cite{Bueno:2003ei} & LAr & 2.4 & Italy & 200 & 200x \\
Daya Bay\cite{Guo:2006ap} & C$_n$H$_{2n}$ & 0.3 & China & 100  & Proposed \\
SNO+\cite{Chen:2005yi} & C$_n$H$_{2n}$ & 1 & Canada & 300 & Proposed \\
CLEAN\cite{McKinsey:2004rk} & Ne,Ar & 0.01 & Canada/USA? & 30 & Proposed \\
HALO\cite{halo} & Pb & 0.1 & Canada & 40 & Proposed \\
MOON\cite{Ejiri:2001ie} & $^{100}$Mo & 0.03 & ? & 20 & Proposed \\
NO$\nu$A\cite{Ayres:2004js} & C$_n$H$_{2n}$ & 20 & USA & 4000 & Proposed \\
OMNIS\cite{Boyd:2002cq} & Pb & 2-3 & USA? & $>$1000 & Proposed \\
LANNDD\cite{Cline:2006st} & LAr & 70 & USA? & 6000 & Proposed \\
MEMPHYS\cite{deBellefon:2006vq} & H$_2$O & 440 & Europe & $>$100,000 & Proposed \\
UNO\cite{Goodman:2001aq} & H$_2$O & 500 & USA & $>$100,000 & Proposed \\
Hyper-K\cite{Nakamura:2003hk} & H$_2$O & 500 & Japan & $>$100,000 & Proposed \\
LENA\cite{MarrodanUndagoitia:2006qs} & C$_n$H$_{2n}$ & 60 & Europe & 18,000 & Proposed \\
HSD\cite{hsd} & C$_n$H$_{2n}$ & 100 & USA & 30,000 & Proposed \\
\br
\end{tabular}
\end{center}
\end{table}

As emphasized above, in order to understand the rates and signatures,
we must understand the nuclear physics involved.  In many cases, the
cross-sections have never been measured experimentally, and
theoretically there are large uncertainties.  One way of decreasing
uncertainties is to use a stopped-pion neutrino source to measure
relevant cross-sections: such a source provides $\nu_\mu$, $\nu_e$, and
$\bar{\nu}_e$ in nearly the same energy range as expected for a
supernova.  A future program of measurements on various targets, such
as that planned for the Spallation Neutron
Source\cite{Efremenko:2005nf}, will be vital for extracting physics
from the next supernova.

\section{Beyond the Milky Way}

Even the largest detectors running today are sensitive only to
supernovae within a few hundred kpc, which pretty much covers only our
own Galaxy.  The next nearest large concentration of stars is the
Andromeda galaxy, about 770 kpc away; at this distance, Super-K would
expect only $\sim$1 event.  Unfortunately, the expected rate of Milky
Way supernovae is only a few per century, so if luck is against us,
the wait may well be longer than a typical physicist's
career.  Several next-generation very large detectors have been
proposed which would have supernova neutrino sensitivity: these
include Mton-scale water detectors (Hyper-K\cite{Nakamura:2003hk}, 
UNO\cite{Goodman:2001aq}, MEMPHYS\cite{deBellefon:2006vq}) 
and 100 kton-scale large LAr and scintillator
detectors (LANDDD\cite{Cline:2006st}, LENA\cite{MarrodanUndagoitia:2006qs}).\footnote{One might consider siting these detectors
to optimize the probability of Earth shadowing\cite{Mirizzi:2006xx}.}
However even the largest of these mega-detectors would see only
tens of events from a core collapse in Andromeda.

But while $1/D^2$ hurts, the increase of potential sources
as $D^3$ helps: a recent study\cite{Ando:2005ka} has pointed out
a regime for which
the probability of detecting a few
events per supernova in a Mton detector is reasonably close to 1 at the
same time as the overall rate of expected core collapses
is reasonably close to 1 per
year.  So if one can operate a large, low background detector
(possibly using optical or gravitational wave (GW) detections nearby in
time to reduce the background), one can expect to collect a thin but steady
dribble of supernova neutrinos.

We can look even farther out: stellar cores have been
giving up their binding energy to neutrinos ever since the first
stars formed, and we are awash in a sea of these ancient neutrinos.
This diffuse supernova neutrino background (DSNB) (formerly known as
the ``relic'' supernova neutrino background) provides a 
steady source of neutrinos.  But because there is no hope of tagging
DSNB neutrinos with optical or GW events, detection feasibility
rests on reducing background to essentially zero.  This may indeed be
possible for $\bar{\nu}_e$ using a large scintillator or Gd-spiked
water detector to tag $\bar{\nu}_e$, in the few tens of MeV regime,
which is nearly free of solar or atmospheric neutrino background.  Detection of
DSNB neutrinos is very interesting from the point of view of
learning about cosmology via knowledge of the past supernova rate.  However,
when considering use of the DSNB to learn about neutrinos, stellar
collapse physics and so on, one must consider the overall rate. One
expects a low, but sure return on one's investment at $\sim$0.1
event/kton/yr of DSNB.  In contrast, in the very long term, on average one
expects about 10 events/kton/yr of Galactic supernova neutrinos.
Counting on a signal from the latter is risky in the short term, because
there may be large Poissonian gaps. But surely, over centuries, the Galactic
supernova detection approach 
wins.  Clearly the best strategy is diversification of one's
experimental portfolio: a large, clean detector that runs for decades
will yield rich and reliable returns.

\section{Conclusion}

Several supernova-neutrino detectors are running and ready to observe
a galactic burst.  A variety of new detectors are proposed: those with
broad flavor sensitivity and tagging ability will be especially
valuable for extracting physics from the signal.  The neutrinos will
come.  We need to build detectors to gather them all: the 
Galactic bursts, the fainter flashes from just beyond, and the dim but
steady background glow.\\

\end{document}